\documentclass[11pt]{article}

\usepackage[preprint]{neurips_2023}

\usepackage{graphicx}
\usepackage{placeins}
\usepackage{float}
\usepackage{subfigure}
\usepackage{booktabs}
\usepackage{natbib}

\usepackage[utf8]{inputenc} 
\usepackage[T1]{fontenc}    
\usepackage{hyperref}       
\usepackage{url}            
\usepackage{booktabs}       
\usepackage{amsfonts}       
\usepackage{nicefrac}       
\usepackage{microtype}      
\usepackage{xcolor}         

\usepackage{multirow}

\begin{document}

\title{Human Activity Recognition Using Self-Supervised Representations of Wearable Data}

\author{
Maximilien Burq \thanks{Equal contributions}\\
Verily Life Sciences  \\
\texttt{mburq@verily.com} \\
\AND
Niranjan Sridhar $^*$\\
Verily Life Sciences \\
\texttt{nirsd@verily.com} \\
}


\date{April 2023}

\maketitle
\begin{abstract}
Automated and accurate human activity recognition (HAR) using body-worn sensors enables practical and cost efficient remote monitoring of Activity of Daily Living (ADL), which are shown to provide clinical insights across multiple therapeutic areas.
Development of accurate algorithms for human activity recognition (HAR) is hindered by the lack of large real-world labeled datasets. Furthermore, algorithms seldom work beyond the specific sensor on which they are prototyped, prompting debate about whether accelerometer-based HAR is even possible \citep{Tong20} .

Here we develop a 6-class HAR model with strong performance when evaluated on real-world datasets not seen during training. Our model is based on a frozen self-supervised representation learned on a large unlabeled dataset, combined with a shallow multi-layer perceptron with temporal smoothing. The model obtains in-dataset state-of-the art performance on the Capture24 dataset ($\kappa = 0.86$). Out-of-distribution (OOD) performance is $\kappa = 0.7$, with both the representation and the perceptron models being trained on data from a different sensor.

This work represents a key step towards device-agnostic HAR models, which can help contribute to increased standardization of model evaluation in the HAR field.



\end{abstract}

\maketitle

\section{Introduction}

Assessments of activities of daily living (ADL) are used in the early detection, diagnosis, treatment, and care of many health conditions, including arthritis, depression, stroke, heart failure, Parkinson's disease, dementia, and Alzheimer's disease \citep{Marshall12, Clynes19, Kazama11, Lawrence14, Fauth13, Sandberg01, s21062047, wainberg2021association, lam2021using}.
Numerous clinical diagnostic tests \citep{UPDRS03, Kroenke01, Rotrou12} include ADL assessments, typically using self-reported data. While such data provide valuable perspectives, complementing them with objective measures can enable comparison of symptoms and outcomes over time or across different patients and allow the effects of interventions to be quantified.
Recent advances in wearable technology have enabled the large scale deployment of sensors and algorithms for human activity recognition (HAR). Such measures allow consumers to track their health metrics over time and are emerging as valuable digital biomarkers in clinical studies \citep{Marra20}.




\subsection{Related work}
Attempts to develop HAR models on smartwatches, smartphones, other body-worn sensors, or via radar, video, and other ambient sensors, have been hampered by a lack of large datasets - ones with high-quality reference activity labels in real-life settings - that can be used to build and test models. For real-world data collection, obtaining high-quality labels of human activities alongside the sensor data stream requires manual annotation of video-based ground-truth, which is costly and can be intrusive for the participants. Alternatively, participants can self-report their activities, but this introduces variability and noise due to participant subjectivity, as well as a selection bias of the activities they choose to report.

\subsection{Supervised HAR models}

Due to these obstacles, most HAR benchmark datasets \citep{Reiss12, Banos15, Bruno13, Altun10, Malekzadeh18, Zhang12} contain data on fewer than 20 research participants, performing 1 to 15 activities for up to 5 minutes under constrained or clinic-like conditions. Given the small size and low ecological validity of these datasets, models trained and tested on them may fail to generalize to the real world when faced with activities not previously encountered, or when applied to new devices or sensors. Furthermore, due to the small size of the datasets available, researchers typically rely on extensive manual feature engineering, which can be lead to overfitting when evaluation is through cross-validation without an explicit test dataset.

While some published studies  report near 100\% accuracy on these datasets individually, cross-dataset performance has not approached such levels of accuracy, prompting debate about whether accelerometer-based HAR is even possible \citep{Tong20}.
The generalization problem strongly constrains the utility of supervised HAR models in practical applications. HAR models trained on small datasets of curated activity labels can only be applied to predict the labels that were present in the training dataset. In real-life data, the vast majority of motion is unconstrained and can easily be misclassified by such models.
Therefore, novel methods for building and evaluating robust and practical HAR models are needed.

One step in this direction is the work by \citep{willetts2018statistical}, and associated open-sourced Capture24 dataset which contains 24h of real-world data from $N=154$ research participants, alongside labels annotated using a body-worn video camera as ground truth. Using a combination of manual feature engineering, tree ensembles and temporal HMMs, they are able to build an accurate ($\kappa = 0.81$) 6-class HAR model.

\subsection{Self-supervised HAR models}

In recent years, self-supervised learning has gained significant attention due to its ability to learn without explicit labels. It involves a pre-training step where a proxy task is defined using information from the data itself. The output of the pre-training step is an encoder which transforms input data into features which are predictive of the proxy task. These features can be used by a new, usually much smaller, model which can be trained or fine-tuned on a labeled dataset to predict a downstream task of interest.
Experiments in the field of image classification have shown \cite{chen2020simple, he2020momentum, chen2020improved} that self-supervised models that have been pre-trained on large unlabeled datasets and fine-tuned on small labeled datasets can match or improve upon fully supervised models which need orders of magnitude more labeled data. This is valuable because, in contrast with labeled data, unlabeled data is usually cheap and abundant.
There is also evidence that self-supervised models can be more robust and generalize better to new datasets \citep{Hendrycks19} than fully supervised models.

Recent attempts to apply self-supervised representation learning to the context of sensor-based Human Activity Recognition have by-and-large focused on demonstrating the generalizability of the \emph{representation itself} by allowing the final few layers of the model to be learned directly on the dataset of interest, often the same small-scale or in-clinic datasets described above \citep{Haresamudram21,Haresamudram22,Saeed_2019}. By testing the models in this way, they typically focus on label efficiency, as a way to assess the representation quality: if the representation is rich enough, the final layers of the model can be trained using few labeled examples. 
\footnote{An important clarification should be made for self-supervised encoder models that are pre-trained on unlabeled datasets and fine-tuned on labeled datasets. Since neural networks can easily memorize small datasets, such evaluations should not be considered evidence for generalization.
Thus in this work, we evaluate our frozen pre-trained representation across multiple datasets and our end-to-end trained model on an entirely new dataset.
}

This approach highlights that self-supervised learned representation are a viable alternative to hand-crafted engineered features as the basis for HAR models. It does not however, address the issue of the end-to-end model generalization to real-world settings.

These past efforts on self-supervised HAR have tried many combinations of optimization objectives, datasets and model architectures for both the pre-training and fine-tuning. Common optimization objectives include enforcing representation invariance across time, augmentations \citep{Haresamudram21,Haresamudram22,Saeed_2019}. Common encoder architectures used include CNNs, RNNs, transformers. Linear or logistic regression models are used for evaluating representation quality while multiple dense layers are often used to make high performance end-to-end models. In \cite{Saeed_2019}, data augmentations are applied on IMU data and a self-supervised is trained to predict the transformation. \cite{Haresamudram22} present an excellent analysis of auto-regressive models and transformers used to learn a representation of the IMU data that can predict neighbouring or future IMU data points, thus capturing the temporal characteristics of the signal. In this work, we combine temporal and transformation constraints to train a robust self-supervised encoder.

Concurrent with this work, \cite{yuan2022self} obtained good results using self-supervised representation as the basis for an HAR model, and evaluate it on the Capture24 dataset ($\kappa = 0.726$) . Although performance is lower than previously published (fully supervised) models on that dataset, this is a promising step towards building a real-world HAR model using self-supervised learning.

    
\subsection{Main contributions}
This works builds on these past efforts and provides three novel contributions:
\begin{itemize}
    \item First, we show \emph{out-of-distribution generalization} $\kappa = 0.7$ of our end-to-end HAR model (including the self-supervised representation) by using the real-world labeled Capture24 dataset for \emph{model evaluation only}. We think that this model evaluation framework is closest to how HAR algorithms are actually used: in real-world settings, and on devices different from the one used for training.
    \item Second, we leverage the self-supervised representations to combine disparate datasets using different sensors: the public in-clinic PAMAP2, together with an internal real-world dataset where labels are obtained from participant-reported tags. Further, we show that the performance of the final model largely improves upon that of models trained on the underlying separate datasets.
    This shows that the self-supervised representations can provide a natural standardization layer and can help reduce the data fragmentation in the field.
    \item Finally, the establish new state-of-the-art performance ($\kappa = 0.86$) for 6-class Human Activity Recognition on the Capture24 free-living video-labeled dataset. This shows that self-supervised pretraining is advantageous even when it is done on a different device.
\end{itemize}




\section{Methods}

\subsection{Datasets}

\subsubsection{Benchmark HAR datasets}
We used 4 public datasets incorporating labeled activities of daily living:  PAMAP2 \citep{Reiss12}, Capture24 \citep{willetts2018statistical}, MHealth and Daily Sports. PAMAP2 contains data of 9 of participants who perform a predetermined set of activities, along with the start and end time of each activity. PAMAP2, MHealth and Daily Sports contain IMU data measured from  wrist-worn devices and are available at the UCI Machine Learning Repository http://archive.ics.uci.edu/ml/index.php.
Capture 24 contains data from 154 participants, followed over 24 hours of unscripted daily living. In addition to the wrist-worn accelerometer, participants wear a body-worn camera which is used by human annotators to generate the activity labels.
Table \ref{tab:bmmeta} summarizes the metadata of the benchmark datasets.


\begin{table*}[htbp]
  \centering
  \begin{tabular}{lccllll}
    \toprule
    Dataset & \#labeled classes & N & Device (sampling rate) & Location & Context \\
    \midrule
    MHealth & 12 & 10 & Shimmer2 (50 Hz) & right wrist & scripted \\
    PAMAP2 & 18 & 9 & Colibri (100Hz) & dominant wrist & scripted \\
    DailySports & 19 & 8 & Xsens MTx (25Hz) & left, right wrist & scripted \\
    Capture24 & 6 & 154 & Axivity AX3 (100Hz) & dominant wrist & real-world \\
    Internal Pilot & 8 & 85 & Verily Study Watch (30Hz) & randomized & real-world \\
    \bottomrule
  \end{tabular}
    \caption{Summary of benchmark datasets metadata.}
  \label{tab:bmmeta}

\end{table*}


\subsubsection{Project Baseline Health Study}
The self-supervised model was pre-trained using a 1-month period of free-living data collected in the Project Baseline Health Study (PBHS) \citep{arges2020project} via a smartwatch (Verily Study Watch) equipped with an accelerometer. 

The Project Baseline Health Study is a prospective, multi-center, longitudinal study that aims to map human health through a comprehensive understanding of the health of an individual and how it relates to the broader population. As part of the PBHS, participants agree to wear a wrist-worn device during daily activities that continuously collects high-resolution IMU data. In our study, we used these data to create a representation of activity without any labels. 

No filtering or inclusion/exclusion criteria were applied beyond those of the main PBHS study, yielding 42,000 hours ($\approx$15 million 10s windows) of accelerometer data from approximately 1200 study participants. During model training, no labels of activity or health were used and the only context used to identify accelerometer data was the device identifier and timestamp. 

\subsubsection{Free-living internal Pilot dataset}
A total of 85 participants were asked to wear the Verily Study Watch over a one to two week period. They were able to self-report activities as they occured throughout their daily living by tagging start and end times directly on the watch. Half of the participants were randomly picked and asked to wear the watch on the right hand and the other half were asked to wear it on the left hand. Participants were instructed to use the firmware on the watch to tag when they were starting an activity and end the tag when they ended the activity. Participants were not asked to record every instance of every activity, but to voluntarily tag as many as possible. We received over 1000 hours of tagged data in 5 activity classes as detailed in Table \ref{tab:fllabels}.

Later, a separate data collection effort was conducted, asking participants to tag the time when they went to bed with the intent of sleeping. In order to more precisely capture sleep epochs, the first and last 15 minutes of each tagged epoch were removed from the final dataset.

\begin{table*}[htbp]
\centering

  \begin{tabular}{ccc}
    \toprule
    Activity & \#hours & \#unique participants \\
    \midrule
    Exercise & 61 & 47 \\
    Sedentary (Sit/Stand) & 420 & 80\\
    In motor vehicle & 175 & 69\\
    Walk/Run & 45 & 36 \\
    Household chores & 134 & 73 \\
    Sleep / in-bed & 196 & 13 \\
    \bottomrule
  \end{tabular}
  \caption{Distribution of activities in pilot dataset.}
  \label{tab:fllabels}
\end{table*}

\subsection{Data preprocessing}
Since the datasets we used have data from different devices and different sampling rates, we first standardized input data by normalizing all values to units of $G$ ($9.8 m/s^2$) and resampled all time series to a regular time-grid at 30Hz. Resampling with and without anti-aliasing did not result in any notable difference in performance. A literature survey shows that most activities of daily living are captured under a frequency limit of 5 Hz, therefore a sampling rate of 30Hz is above the Nyquist frequency and aliasing should not be a concern. Next, we divided the time series into non-overlapping windows of 10 seconds each.

Each window contains 3 time series corresponding to $x, y$ and $z$ components of IMU identified by the device identifier and the starting timestamp of the window.
Although the various devices are not always guaranteed to have the x, y and z channels map to the same axes in the watch reference frame, we did not standardize them prior to training. Because of natural variations of watch wear on the left or right wrist, along with rotations around the wrists, these x, y and z channels cannot be assumed to be fixed in wrist-frame even for a given device type. Therefore, we elected to handle these sources of variability through data augmentations directly when training the representation model.

\subsection{Self-supervision labels}
\begin{figure}[htbp]
  \centering
  \includegraphics[width=\linewidth]{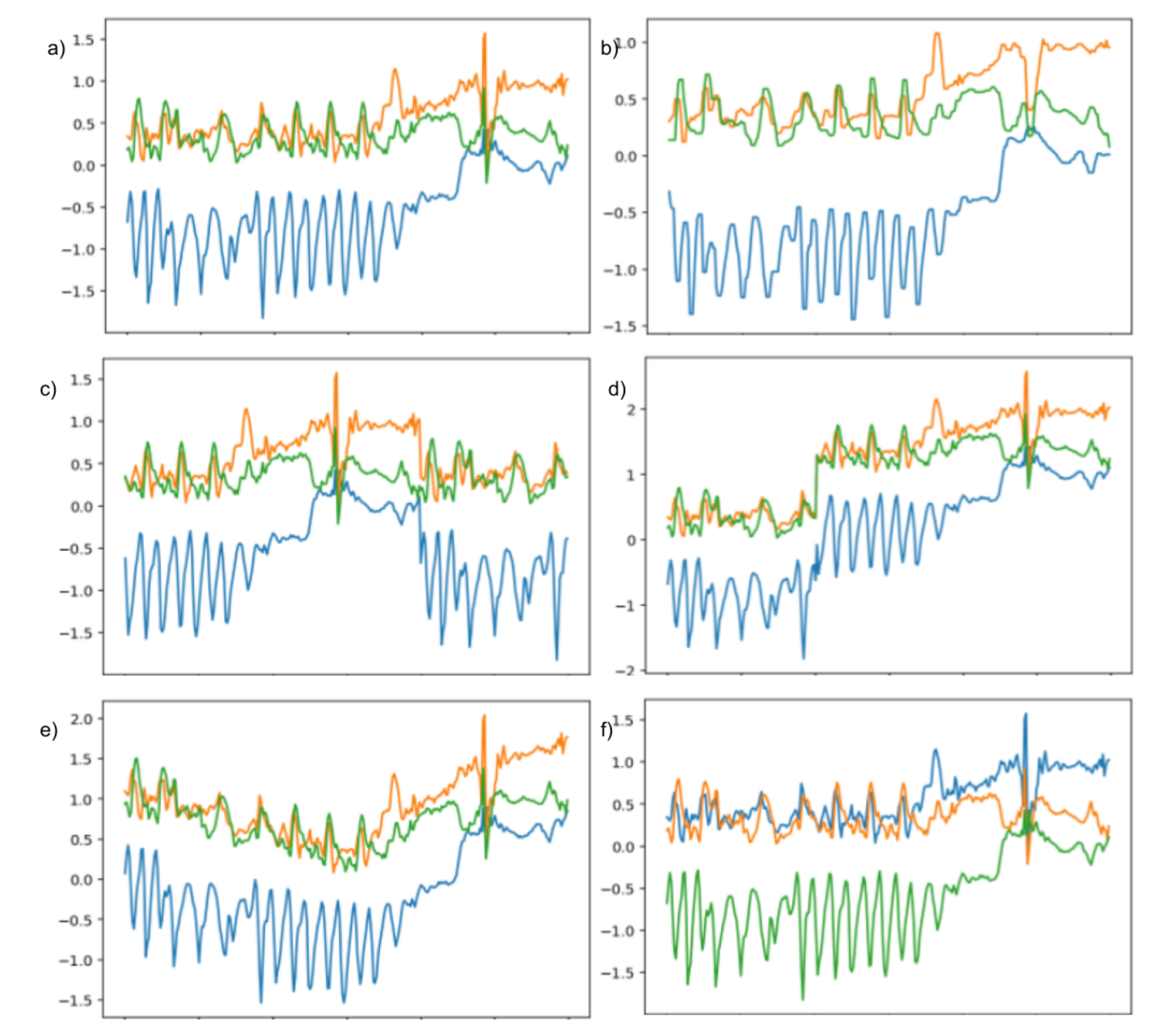}
  \caption{Augmentations. a) is the original window and the rest are its augmented versions. b) is the smoothed window, c) has been translated in time, d) has a discontinuous jump in the baseline, e) has a gradual baseline wander, and f) is rotated in 3-axis.}
  \label{fig:augmentations}
\end{figure}

We use a contrastive approach based on temporal coincidence for self-supervised learning. The principle is to learn a mapping from input (in this case, windows of accelerometer data) to an embedding vector space in which coincident pairs are closer to each other than pairs which are not coincident. Thus the choice of coincidence criteria decides the geometrical relationship between points in the embedding space. In this work, 2 main types of coincidence were used:
\begin{enumerate}
    \item \emph{Temporal proximity}: two windows were considered as coincident if they belong to the same user and occur within a specified time $\Delta t$. The temporal proximity criterion creates clusters of ‘slow-moving’ activities (i.e. activities that have the same underlying structure over periods as long as or longer than $\Delta t$). For this study, we chose a maximum temporal proximity distance $\Delta t$ = 60 seconds.
    \item \emph{Augmentation}: a transformation was applied on the accelerometer signal and the original and augmented views were considered coincident. This criterion makes the representation invariant to certain perturbations that resemble common forms of noise, or variations seen within and between accelerometers. For example, we added random jump in baselines, baseline wander, median filtering, rotation in the x-y plane and Gaussian noise. For each of these augmentations and their compositions, we made pairs using the original window and its augmented version. Figure \ref{fig:augmentations} shows the transformations we used, along with examples of their effect on the data.
\end{enumerate}

The construction of the optimization objective was inspired by the recent work in images and audio \citep{Jansen19, Shaoqing18}. Coincident pairs are created before training by sampling data from the full dataset and applying the constraints described above. Thus the input to the encoder with a batch size of $b$ contains $b$ pairs of IMU windows where each pair is randomly sampled from the dataset.

\begin{figure}[htbp]
  \centering
  \includegraphics[trim={1cm 3cm 2cm 2cm},clip,width=\linewidth]{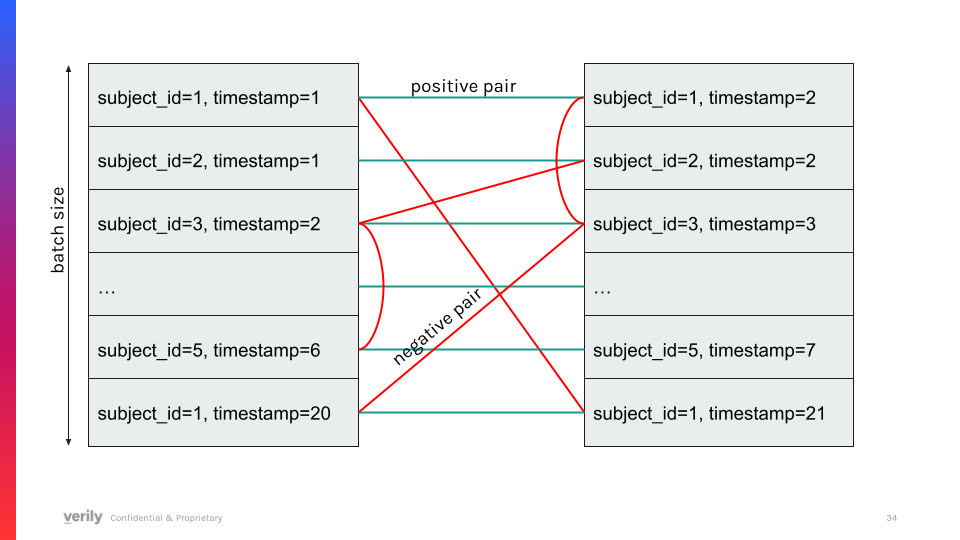}
  \caption{Pair labeling network: Training inputs are batches of paired windows. Pairs $ii$ (connected by green lines) for $i^{th}$ element of the batch are positive pairs that were sampled before training. Pairs $ij$ (connected by red lines) for $i != j$ are random combinations. For a large shuffled dataset, these combinations can be assumed to be negative pairs.}
  \label{fig:batchlabel}
\end{figure}

Figure \ref{fig:batchlabel} depicts how negative pairs were created during training, we simply consider all the possible pairs that can be constructed by $2b$ IMU windows: $4b^2$. There are $b$ ($*2$ permutations) pairs with positive labels; these were the coincident pairs created in the previous step. We have an additional $2b$ identity pairs (pairs of the window with itself). The remaining $4b^2 - 4b$ pairs are combinations of windows which are randomly sampled from the dataset and can be considered unrelated pairs and thus labeled negative. This is a valid approximation that holds for large datasets where the number of subjects $> b$ and the number of windows per subject $>>b$. Since there were many more negative pairs than positive, we down-weighted negative labels by a factor of $(2b-2)$ and down-weighted the identity pairs to 0. 

\subsection{Pre-training}

\begin{figure}[htbp]
     \centering
     \begin{subfigure}
         \centering
         \includegraphics[trim={1cm 1cm 0 0},clip,width=0.9\linewidth]{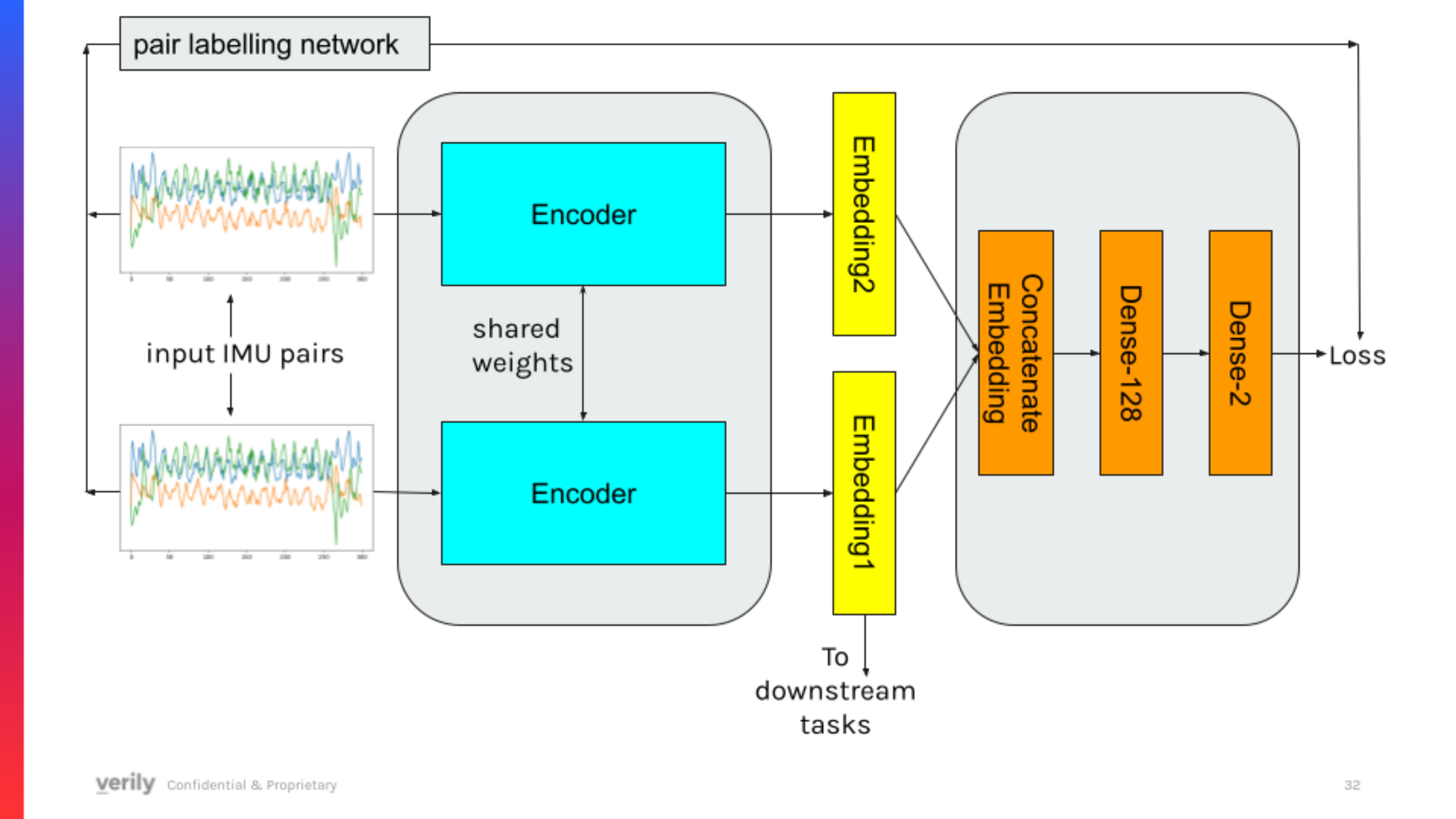}
     \end{subfigure}
     \hfill
     \centering
     \begin{subfigure}
         \centering
         \includegraphics[trim={1cm 1cm 0 0},clip,width=0.9\linewidth]{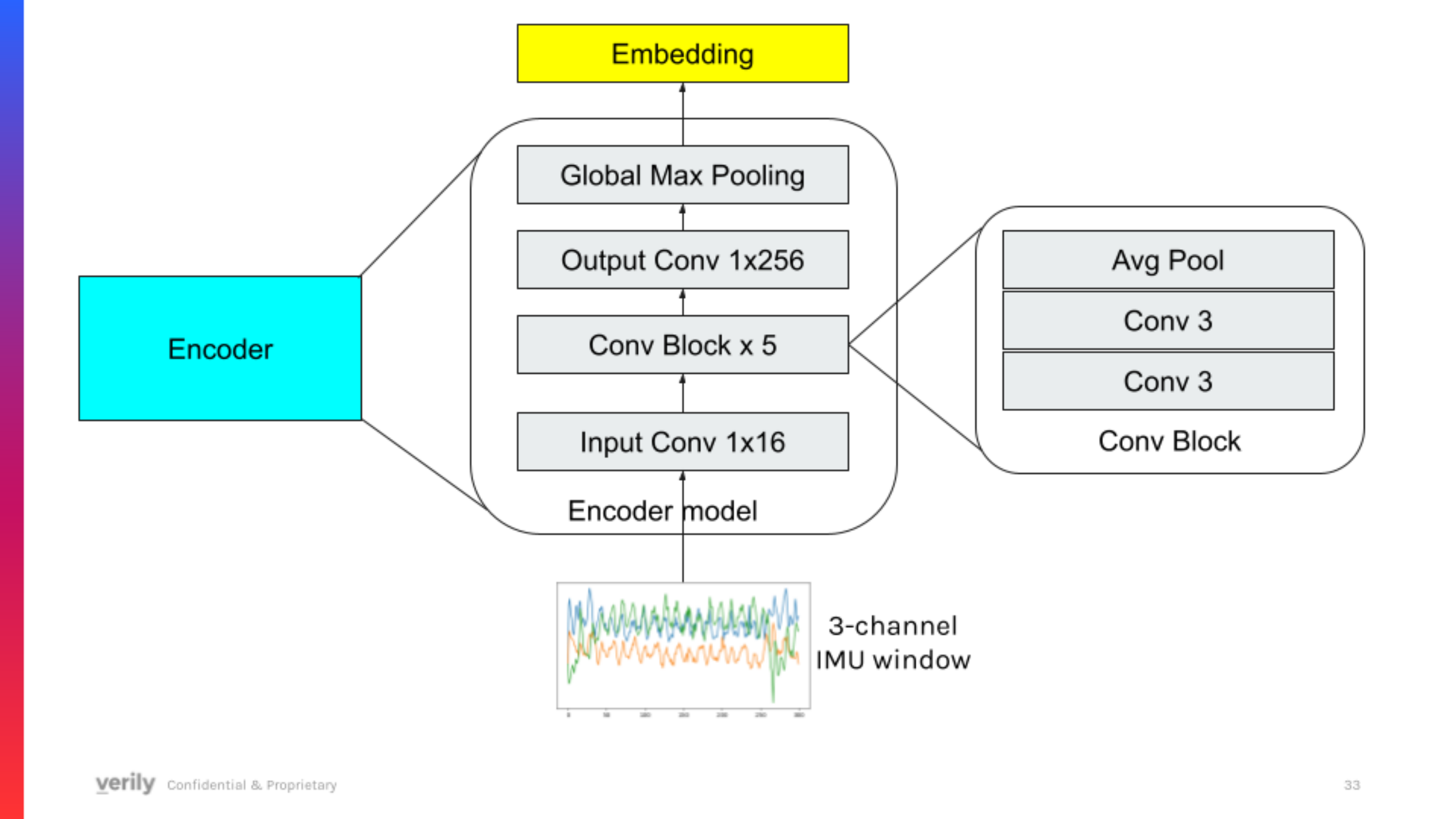}
     \end{subfigure}
     \hfill
\caption{Model training pipeline. The main model is the convolutional neural network encoder which takes 10 second accelerometer windows as input and returns a vector embedding. The auxiliary dense layers are added during training to predict which pairs of IMU windows are labeled as positive.}
\label{fig:trainpipeline}
\end{figure}

The encoder ‘encodes’ a given 10-second window of 3-axis accelerometer data into a 256-dimensional feature vector. The encoder architecture of the model is a convolutional tower with c convolutional blocks, followed by a dense layer with d-dimensional embedding output. We experimented with various hyperparameters and achieved stable results with $c=5$ convolutional blocks and the $d=256$ embedding size. 

During training, coincident pairs are picked randomly from the dataset, stacked in batch dimension and fed to the encoder to get embeddings for all windows (i.e., if the original batch size is $b=128$, effective batch size to encoder is $2*b = 256$). We then created paired embeddings by concatenating the embeddings of all possible pairs of windows in the batch (i.e. $(2*b)*(2*b) = 65,536$ paired embeddings). The resulting paired embeddings were then fed into a 2-layer perceptron projector. The output of the perceptron is a binary softmax prediction, with negative or class 0 indicating the two windows are not coincident and positive or class 1 indicating that the two windows are coincident. The corresponding positive and negative labels are created as described above. The training loss was a cross-entropy loss between the predicted binary prediction and the coincidence labels. This loss - which uses only the IMU windows and the device and timestamp identifiers of the windows and no external oracle and thus \emph{self-supervised} - cause the encoder to extract features that are invariant to noise and can characterize motion that is consistent over time. Training was performed for 0.5 million steps and the model was then saved. 

During inference, we drop the loss network, and the encoder takes as input a single 10-second window and returns a 256-element feature vector or embedding. Figure \ref{fig:trainpipeline} shows the architecture of the full training pipeline, specifically the encoder model.

\subsection{Training the final HAR head}

Some of the evaluation datasets may have severe class imbalance. This is especially the case for the ones where data is collected in the real world, and where class balances are reflective of the natural occurence of various activities. In order to avoid class collapse when training with categorical cross-entropy, we re-weight the minority class to be no less than 5x smaller than the majority class. During evaluation, the original class balance is used.

Our model is a simple fully-connected perceptron with 5 layers and 128 neurons each. At inference time, class prediction logits are smoothed temporally over a 2-minute window. We find that this temporal smoothing mechanism increases classification accuracy.


\subsection{Benchmark}
In order to better understand the benefit of self-supervision, we provide performance when a simplified HAR head (1-layer perceptron) is trained using only 8 first and second-order statistics of the 10s signal: the means and std of the acceleration along the xyz axes, and the mean and standard deviation of the vector norm of the acceleration. We find that these features, termed \emph{benchmark} yields surprisingly strong results which outperform some of the existing published results on our datasets of interest.

\section{Results}

\subsection{Accuracy of self-supervised representations}
Similar to previous work, we first evaluate the usefulness of the representation model by training a simple linear classification  model on top of the frozen representations.
In Table \ref{tab:accuracy}, we show that our representations are useful for human activity recognition, leading to strong classification accuracy\footnote{Because we focus on generalization beyond any single dataset, here we provide comparisons only with previous work where model performance is provided for at least two distinct datasets.}, and can work across multiple sensors and datasets.

\begin{table}[htbp]
\centering
  \begin{tabular}{cccc}
    \toprule
    Authors (year) & PAMAP2 & MHealth & DailySports  \\
\midrule
J. Wang et. al (2018)  & 39.21 & - & 57.97  \\
Xin Qin et al (2020)** & 63.9 & - & 60.7 \\
Holzemann et al (2020) & 20.1 & - & -\\
TransAct (2017) ** &  - & 82 & 85 \\
\midrule
Ours (benchmark) & 74.3 & 82.4 & 72.8\\
Ours (self-supervised) & \textbf{83.3} & \textbf{93.4} & \textbf{91.1}\\
    \bottomrule
  \end{tabular}
  \caption{Random test set activity classification accuracies achieved by comparable efforts that show generalization of models. ** indicates the performance is measured on a smaller set of activities than what is found in the dataset.}
\label{tab:accuracy}
\end{table}

When evaluating models on real-world datasets such as Capture24, class imbalance can become a primary concern. As a result, previous works often report on $\kappa$ scores rather than raw accuracy. In Table \ref{tab:kappas}, we provide these numbers for the Capture24 dataset.

\begin{table}[htbp]
\centering
  \begin{tabular}{ccc}
    \toprule
    Authors (year)  & Capture 24 \\
\midrule
\cite{willetts2018statistical}  & .81 \\
\cite{yuan2022self}    & .726 \\
Ours (self-supervised) [2023]  & \textbf{.86} \\
    \bottomrule
  \end{tabular}
  \caption{$\kappa$ scores for comparable efforts at building general models}
    \label{tab:kappas}

\end{table}

\subsection{Out-of-distribution generalization}
To test the out-of-distribution generalization of our framework, we train 4 models on top of the self-supervized frozen representations, where the final classification head are trained respectively on Capture24, Pilot and the combination of PAMAP2 and Pilot.
The lack of standardization of label semantics across datasets poses a key challenge arises when attempting to evaluate out-of-distribution generalization for an HAR model.
Table \ref{tab:semantics} shows how the classes of the PAMAP2 and Pilot datasets map to the classes in Capture24.
Although we take special care when mapping the class semantics of one dataset to those of another, some semantic differences remain, and likely account for some of the lower out-of-distribution performance compared to in-distribution.

\begin{table*}[htbp]
\centering
  \begin{tabular}{ccc}
    \toprule
    Capture24 class & PAMAP2 classes & Pilot classes \\
\midrule
Sleep & - & sleep / in-bed \\
\midrule
\multirow{5}{*}{sit-stand}  & stand & standing in place  \\
& sit & still \\
& computer &  \\
& lying & \\
& TV & \\
\midrule
vehicle & drive & in motor vehicle \\
\midrule
\multirow{5}{4em}{walking} & walk & slow walking  \\
& run & Walk-run \\
& desc stairs &  \\
& nordic\_walk &  \\
& asc stairs &  \\
\midrule
\multirow{6}{*}{mixed activity}  & clean house & household chores  \\
& fold\_laundry &  \\
& iron & \\
& rope\_jump &  \\
& soccer &  \\
& vacuum &  \\
\midrule
bicycling & cycling & sports \\
    \bottomrule
  \end{tabular}
  \caption{Model performance when varying the data used for training the final HAR head. * indicates that the training and evaluation datasets are different, indicating out-of-distribution setup.}
    \label{tab:semantics}

\end{table*}


In Table \ref{tab:ood}, we provide the out-of-distribution generalization of various models where the final training is performed on either Capture24, PAMAP2 or the Pilot dataset.

\begin{table*}[htbp]
\centering
  \begin{tabular}{cccc}
    \toprule
    Training data & metric & Eval: Capture24 & Eval: Pilot \\
\midrule
Capture24  & $\kappa$ & .86 & .66* \\
Pilot dataset  & $\kappa$ & .54* & .72 \\
Pilot dataset + PAMAP2  & $\kappa$ & \textbf{.70*} & .71 \\
    \bottomrule
  \end{tabular}
  \caption{Model performance when varying the data used for training the final HAR head. * indicates that the training and evaluation datasets are different, indicating out-of-distribution setup.}
    \label{tab:ood}
\end{table*}

In particular, we observe that the model trained on the combined Pilot + PAMAP2 dataset has similar out-of-distribution performance (on Capture24) and in-distrubution (on the Pilot data), indicating good generalization performance.

\begin{figure}[htbp]
     \centering
         \includegraphics[clip,width=0.49\linewidth]{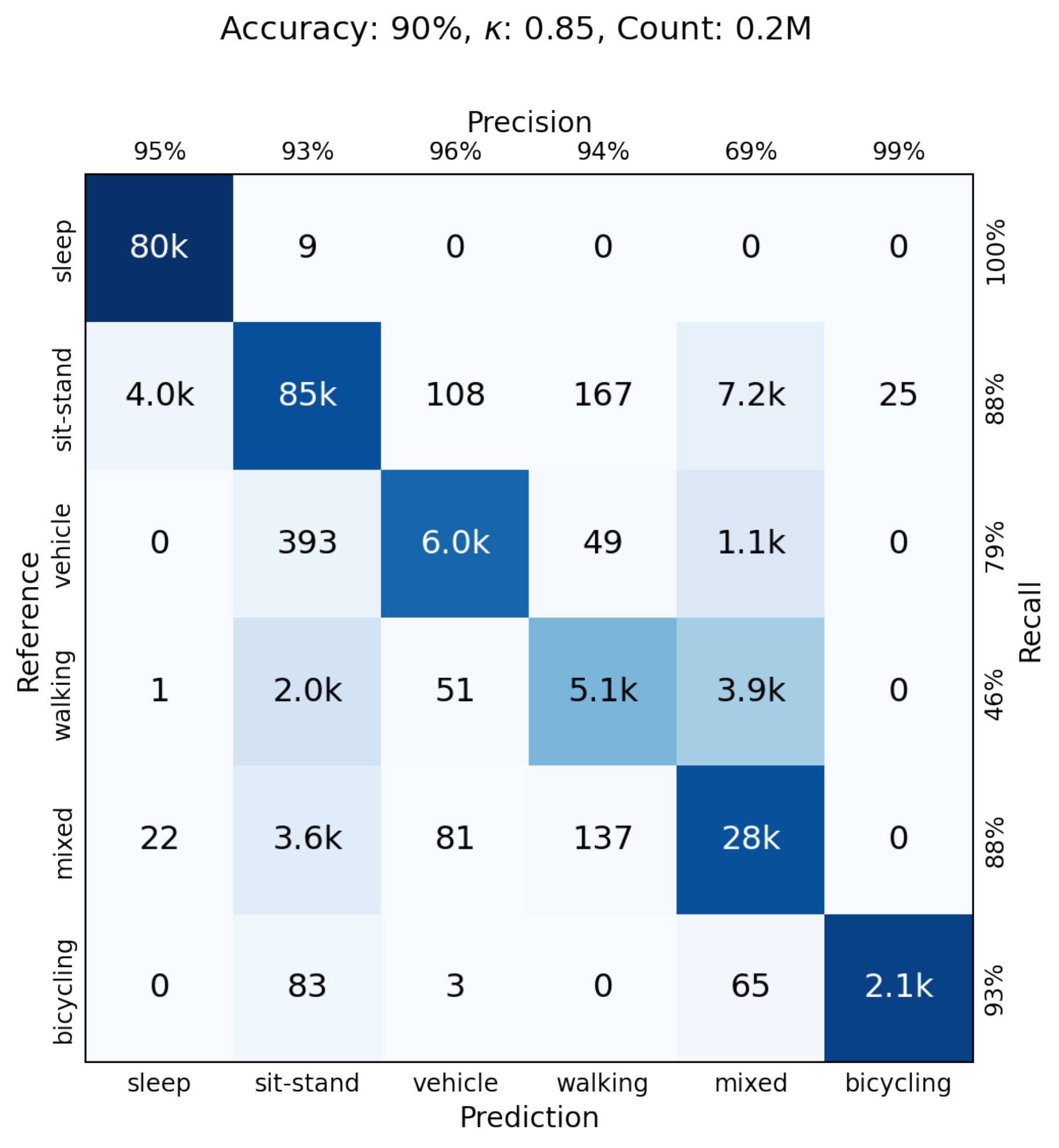}
         \includegraphics[clip,width=0.49\linewidth]{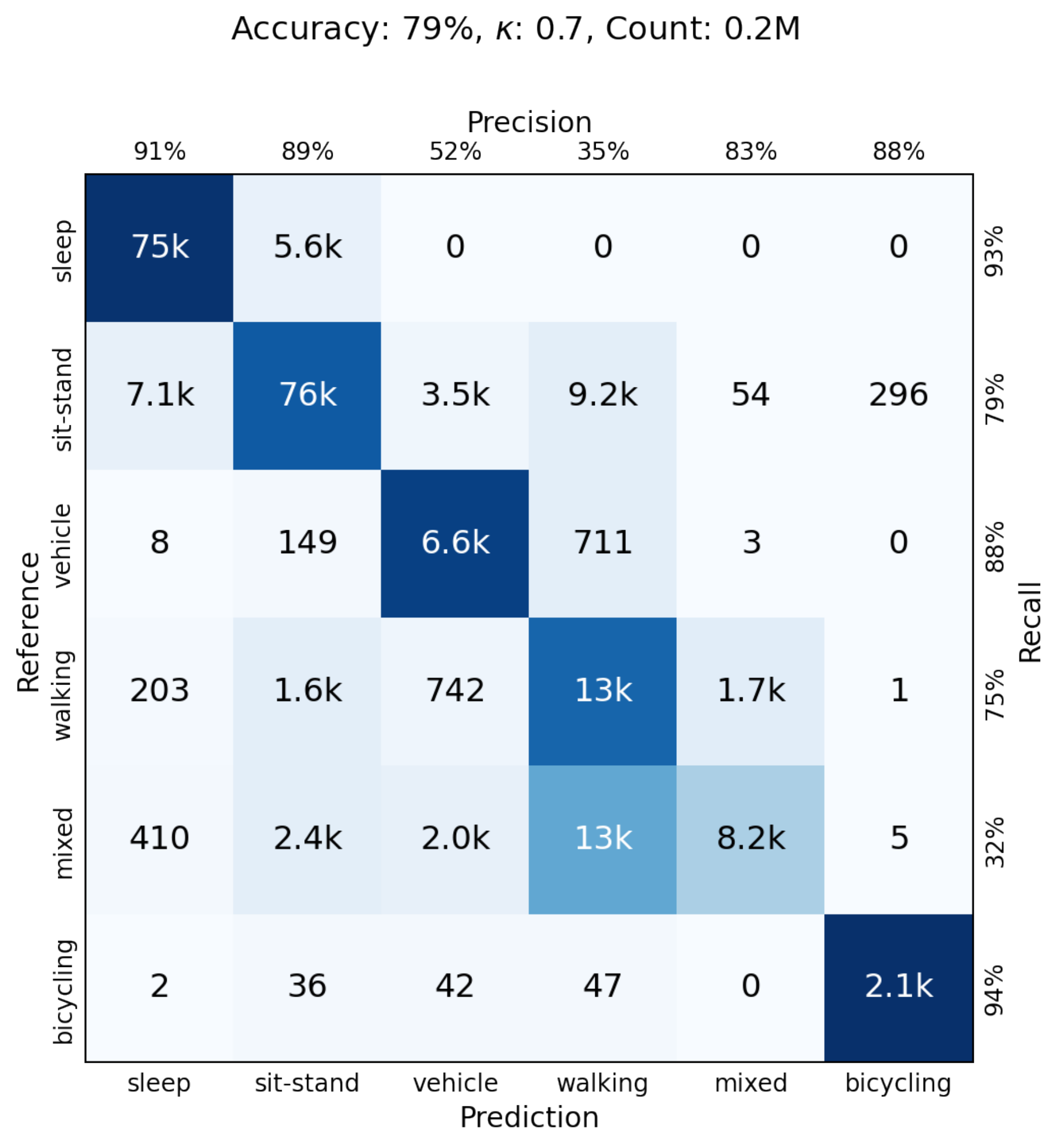}

\caption{Confusion matrices for on the Capture24 dataset. Left is the model trained on Capture24 (in-distribution). Right is the model trained on PAMAP2 + Pilot datasets (out-of-distribution).}
\label{fig:ood}
\end{figure}

In Figure \ref{fig:ood}, we observe the sources of misclassification for both the in-distribution Capture model, as well as our best combined model.

\section{Discussion}
Our work combines some of the documented attributes of previous efforts in HAR, such as self-supervised representations, data augmentations and temporal proximity pairs, and combine that with a large pre-training dataset. 
We demonstrate the power of using a large unlabeled training dataset to pre-train powerful self-supervised representations of human motion. We show that training a CNN on a large dataset can produce state of the art results and show excellent generalization to new datasets and new devices. Using both the temporal proximity and data augmentation loss provide sufficient context for our model to learn highly discriminative features representations.
We demonstrate that the learned representation can extract highly discriminative features which can be used to recognize human activity.

Therefore in a practical application, we believe the essential first-order component of a robust HAR system is a robust feature representation  that can generalize to new devices and subjects.


Finally, we summarize the considerations we implement to ensure evaluation and comparison of HAR models. In particular, we note that our benchmark evaluation shows that even trivial statistical features can easily show very strong performance within a single dataset, therefore it is a useful method to assess the value of complex models and the ease of separability of different datasets. The benchmark models also fail when trained on one dataset and applied on a new dataset, highlighting the importance of testing cross-dataset generalization.

\subsection{Limitations and future work}
There are a few limitations of our proposed HAR framework. First, the training objective of our self-supervised loss assumes that temporally adjacent windows are likely to be part of the same activity. This objective enables the encoder to learn a rich representation for ‘slow-moving’ activities (i.e. activities that have the same underlying structure for at least 1-5 minutes). However, this objective could potentially act against learning rich representations of activities that last less than 1 minute (e.g. standing up from a chair or opening doors). The second training objective, which uses augmentations instead of temporal proximity for defining coincidence pairs, is better suited to represent these activities. More analysis is needed to quantify the impact of the 2 objectives and how they affect different activities. This will require more detailed studies with many more activities than are found in the benchmark datasets.

Our design choice of providing classification at the 10 second level fundamentally limits the time scale at which representations are learned. As a result, this limits the model's ability to recognise more complex activities which may occur at the 5-30 minute time scale. Further work with longer time windows could lead to improvements, and reduce the need for our own approach of using temporal smoothing to increase model robustness.

\begin{ack}
We thank Erin Rainaldi, Ritu Kapur, Anthony Chan, David Andresen and Stephen Lanham for the organization, collection and compilation of the free-living pilot dataset with the Verily Study Watch. We thank Aren Jansen for invaluable guidance and feedback in designing the concept and implementation of  self-supervised machine learning models. We thank Sara Popham for critical feedback and analysis of the model. We thank the UCI Machine learning dataset repository, Mohammad Malekzadeh, and USC for hosting the public benchmark datasets used in this effort. We thank all the participants in the Project Baseline Health Study and the pilot study for their participation.
\end{ack}

Verily Life Sciences, LLC, funded the work and the research studies. All authors were full time employees of Verily Life Sciences during their contributions to this effort. No financial compensation was received outside of the contributors’ regular monetary and stock compensation due to their employment at Verily Life Sciences.

\bibliographystyle{plainnat}
\bibliography{main.bib}


\end{document}